# A novel large-volume Spherical Detector with Proportional Amplification read-out

I. Giomataris[1], I. Irastorza[2], I. Savvidis[3], S. Andriamonje[1], S. Aune[1], M. Chapelier[1], Ph. Charvin[1], P. Colas[1], J. Derre[1], E. Ferrer[1], M. Gros[1], X.F. Navick[1], P. Salin[4], J. D. Vergados[5]

*1 : IRFU, Centre d'études de Saclay, 91191 Gif sur Yvette CEDEX, France*
*2: University of Saragoza, Spain*
*3 : Aristotle University of Thessaloniki, Greece*
*4 : APC, Université Paris 7 Denis Diderot, Paris, France*
*5: University of Ioannina, Greece*

**Abstract**
A new type of radiation detector based on a spherical geometry is presented. The detector consists of a large spherical gas volume with a central electrode forming a radial electric field. Charges deposited in the conversion volume drift to the central sensor where they are amplified and collected. We introduce a small spherical sensor located at the center acting as a proportional amplification structure. It allows high gas gains to be reached and operates in a wide range of gas pressures. Signal development and the absolute amplitude of the response are consistent with predictions. Sub-keV energy threshold with good energy resolution is achieved. This new concept has been proven to operate in a simple and robust way and allows reading large volumes with a single read-out channel. The detector performance presently achieved is already close to fulfill the demands of many challenging projects from low energy neutrino physics to dark matter detection with applications in neutron, alpha and gamma spectroscopy.

## 1. Introduction

There is increasing interest for massive low-background, low-energy threshold detectors in particle astrophysics for seeking the origin of dark matter in our universe and studying low-energy neutrino physics [1]. Large gaseous detectors located underground have been proposed to pursue such important new physics possibilities [2].

The search for WIMP (Weakly Interacting Massive Particles) dark matter at ever increasing levels of sensitivity requires detectors scalable in target mass to 1000 kg and beyond whilst maintaining the ability to reject backgrounds that would simulate a signal for dark matter. High precision gaseous detectors reading out large drift volumes could be used to measure direction of produced recoil during WIMP- nucleus collision [3,4,5].

Another leading dark matter candidate is the axion, a hypothetical elementary particle proposed as a solution to the 'strong CP' problem. Standard solar axions are under study by the CAST experiment at CERN [6] using precise gaseous detectors for keV x-rays among a large background of cosmic rays and environmental radioactivity.

The detection of low energy neutrinos [7] especially via coherent nuclear scattering, arising from the neutral current interaction, remains a great challenge in neutrino physics. At such energies the long neutrino wavelength probes the entire nucleus, giving rise to a large coherent enhancement in the interaction cross section via a neutral current, proportional to neutron number squared. Such an important increase of the cross section is attractive for many applications and especially for the measurement of the reactor or Supernova neutrinos by using relatively 'light' mass detectors [22]. The energy transferred to the recoiling target is however, quite low, a few keV at most, even for the lightest nuclei.

Other potential applications requiring large volume of about 10 m in radius are described in detail in reference [1,8-10]. Low energy neutrinos from strong tritium source could be studied with such massive detector opening the way to determine neutrino oscillation parameters. This detector could also be sensitive to the neutrino magnetic moment and be capable of accurately measure the Weinberg angle at low energy [1].

In addition, for the search of Dark matter in underground area, it is crucial to know with a good precision the contribution of the neutron background [11]. At present a very high efficiency and a very good energy resolution neutron detector able to measure very low neutron flux ($10^{-6}$ n/cm$^2$/s) in a large energy range from thermal to several MeV does not exist. The neutron detector, based on the same geometry, will use Helium-3 on a large mass scale. The resulting good resolution detector could provide the neutron energy spectrum at ground and underground installations. Good energy resolution is also required to obtain a low energy threshold and identify lines from gamma background.

For theses reasons we have developed a new gaseous detector based on a spherical geometry. It combines large mass, sub-keV energy threshold and good energy threshold. A prototype of a spherical detector of 1.3 m diameter has been built, in the last two years at Saclay, as a demonstrator and its initial operation is quite promising. The design presents two basic innovations:
- A large spherical drift vessel.
- A new proportional spherical counter located at the center of the drift vessel.

In view of the very good energy resolution obtained with the preliminary test of the Rn decay, we believe, this type of detector can also be used for the double beta decay using a $^{136}$Xe gas. In the next sections full details on the new instrument and its performance will be provided.

## 2. Detector description

The detector consists of a hollow copper sphere (the sphere is actually an old LEP cavity) of 1.3 meter in diameter and 6 mm thick surrounding a small stainless ball of 1.4 centimeter in diameter. It operates in a seal mode: the spherical vessel is first pumped out and then filled with an appropriate gas at a pressure from low pressure up to 5 bar. The ball is maintained in the centre of the sphere by a stainless steel rod. In the simplest design the small spherical electrode acts as proportional amplification counter and is read by a single channel electronic chain. A picture of detector and surrounding equipment is shown in Figure. 1

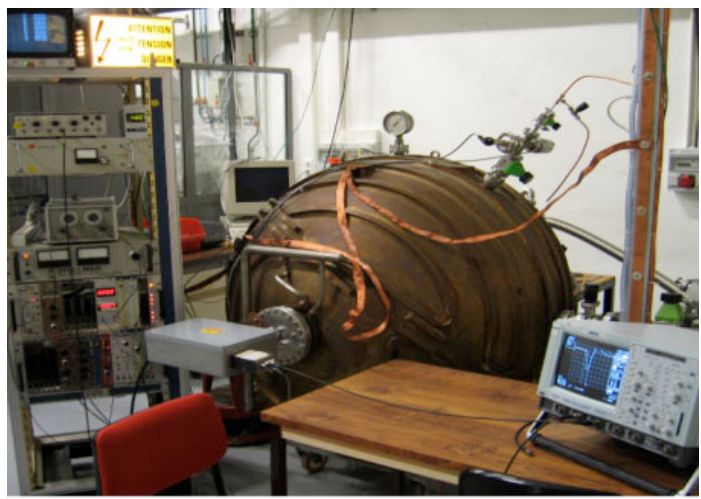

**Figure 1**. A photograph of the spherical vessel and electronics

The detector was operated with positive bias applied to the anode (inner sphere) while the cathode (external sphere) remained at ground potential. A high voltage capacitor was decoupling the high voltage cable to protect the sensitive preamplifier. The anode ball 14 mm in diameter was made out of stainless steel.

In the spherical proportional detector, the electric field depend on the anode radius $r_2$, cathode radius $r_1$, anode voltage $V_0$ and radial distance r:

$$E(r) = \frac{V_0}{r^2} \frac{1}{1/r_2 - 1/r_1} \quad (1)$$

Using $E = -\nabla V$ we can calculate the potential V as function of radius r:

$$V = V_0 \frac{1/r - 1/r_1}{1/r_2 - 1/r_1} \quad (2)$$

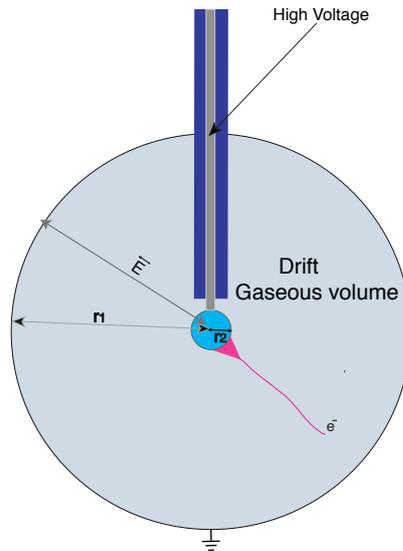

**Figure 2.** A positive High Voltage (HV) applied in the central producing a radial electric field in the inner spherical volume. An electron created in the gas volume drifts to the central electrode producing an avalanche near the spherical ball at a distance of a few mms. Positive ions moving backward are inducing a signal to the preamplifier.

Energetic charged particles, x-rays, or gamma rays or even neutrons entering the detector strip electrons from the gas atoms to produce positively charged ions and negatively charged electrons. The electric field created across the electrodes drifts the electrons to the positive electrode. Near the inner anode sphere the electric field is high enough and electrons gain enough energy to ionize more gas atoms, a process that produces more electrons. Typical gases at atmospheric pressure required field strength on the order 10kV/cm to produce the avalanche of secondary electrons around the small anode ball.

The avalanche is produced at a few mms distance from the anode and the positive ions drifting toward the cathode are inducing a pulse to the charge preamplifier.

Since the avalanche takes place near the small ball and the electrons are attracted to it the positive ions travel a much greater distance. Therefore the induced pulse to the preamplifier is mainly due the ion movement; electrons produced during avalanche process have a negligible contribution to the signal.

By definition, the mobility, µ, of an ion in a gas is the ratio of its drift velocity to electric field and it is roughly constant:

µ = v/E(r) = dr/dt(1/E(r))

Substituting E(r) from equation (1) we find :

$$r^2 dr = \mu \frac{V_0}{1/r_2 - 1/r_1} dt$$

So the ions arrive at distance *r* at the time t given by:

$$\int_{r_2}^{r} u^2 du = \int_{0}^{t} \mu \frac{V_0}{1/r_2 - 1/r_1} du$$

If we substitute this expression of *r(t)* in the equation (2), we get the signal given by the detector for an avalanche created at time *t=0*:

$$V(t) = \frac{V_0}{(1/r_2 - 1/r_1)} \left[ \left( \frac{3\mu V_0 t}{1/r_2 - 1/r_1} + r_2^3 \right)^{-\frac{1}{3}} - 1/r_1 \right] \quad (3)$$

From (3) we can derive the expected shape of the induced signal by taking into account preamplifier shaping constant. We used a low-noise charge amplifier with large RC to pick-up the induced signal. Using the transfer function of the amplifier we get a signal which reproduced quite well the pulse observed from an $^{55}$Fe source as shown in Figure.3. The agreement between calculation (red line) and observation (black line) is rather satisfactory.
More sophisticated options are envisaged for future prototypes readouts, the preferred choice being Micromegas detector [12,13] which provides high precision, fast response and excellent energy resolution.

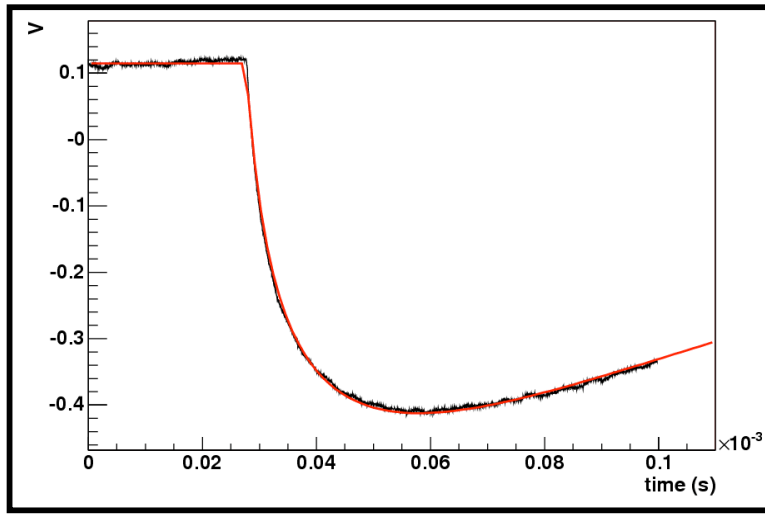

Figure 3. Pulse observed by a 55Fe x-ray (black line) well reproduced by the calculation (red line).

4. **Electrostatics of the Spherical TPC**

In a real implementation of the spherical TPC concept, the ideal spherical symmetry is broken by the rod that supports the central electrode and that necessarily connects it to the front-end electronics, placed outside, to amplify and read the signals. In Figure 4 the equipotential lines are plot for this simplest geometry, showing how the presence of this rod makes the electric field to be far from spherically symmetric. In particular, by doing simple electrostatics calculations, like the one in Figure 5, we see that only about 20 % of the volume have field lines that end up in the spherical electrode, while the rest go to the cylinder, in which no amplification (or a wrong one) is expected.

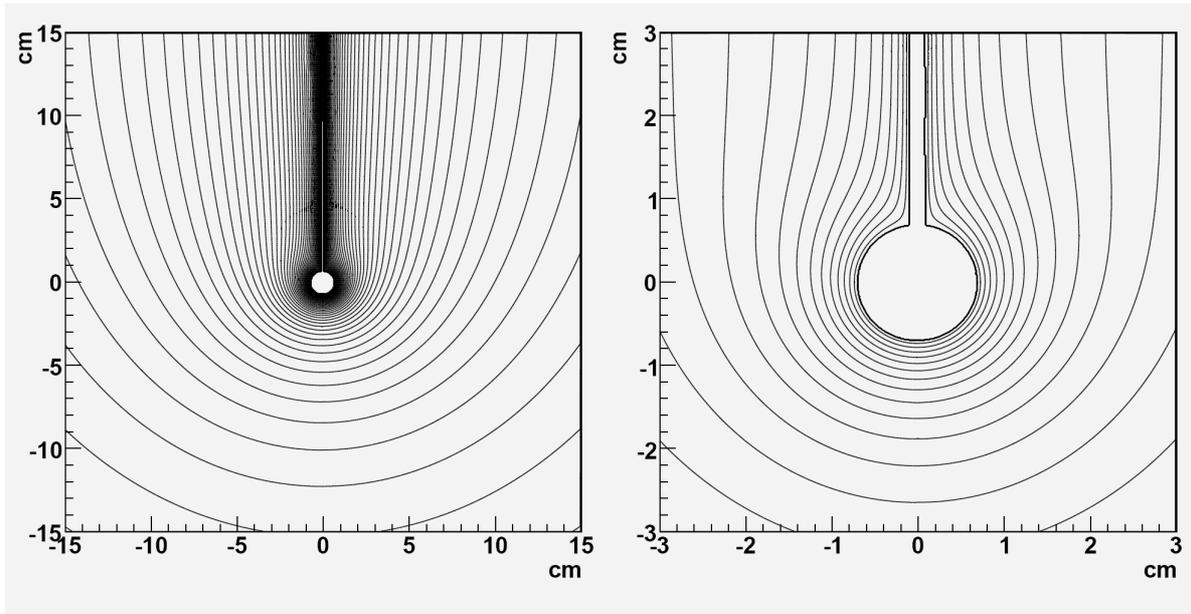

**Figure 4. Electric field (equipotential lines) around the simplest electrode geometry (sphere connected to stick)**

Moreover, a close inspection of the voltage gradient in the region very close to the electrode surface (less than few mm), where the amplification of the signal is expected to occur, shows variations along the surface which leads to amplification factors that will depend on the precise point on the electrode surface where the field line ends. This inhomogeneity in gain is magnified by the usual exponential dependence of the avalanche gain with the electric field. Therefore, this simple geometry, although necessary to provide a proof of the principle of detection with a point-like source as shown in the previous section, is unsuitable for a serious application where a homogeneous response over the detector volume is required.

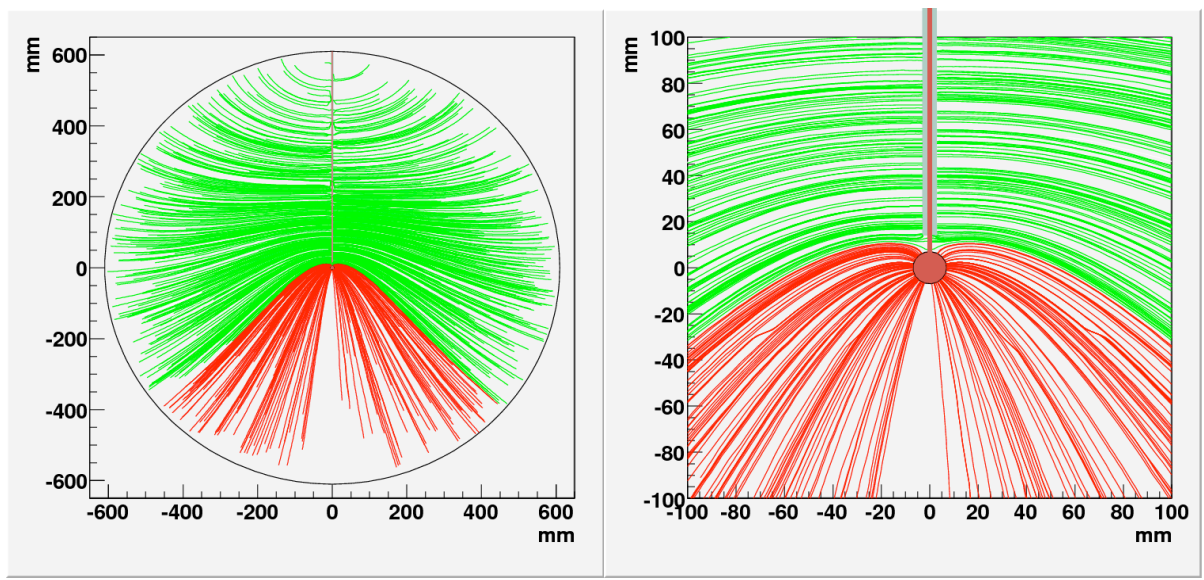

**Figure 5.- Drift field lines for the simplest electrode gemetry (sphere connected to stick). The lines have been calculated starting from 1000 points randomly distributed over the sphere volume. Colors are used to distinguish lines ending on the surface of the spherical electrode (red) from those ending on the stick (green). The fraction of the detector volume which lines end up in the sphere is calculated to be about 20%. Left one shown the overall picutre and right a close-up of the 10 cms around the center.**

Current efforts focus on the design of an electrostatic structure that allows bringing the high voltage to the internal sphere with minimal distortion of the spherical field, both for purposes of drift and homogeneous amplification all around the small sphere. Various ideas of such field correctors have been tested but this will be widely described in the next section.

## 5. Results in the simplest electric field configuration

In this section we will give results obtained in the simple mode without using any field correction system but simply testing only the 'good' hemisphere. In this compartment a 55Fe source is introduced inside the sphere by means of a movable stick which allows us to place the source at any distance from the inner electrode

The first tests were oriented to the assessment of the tightness of the vessel, so the gas could keep the sufficient level of purity for right operation. The volume was pumped by a primary pump followed by a turbo molecular pump, reaching a level of vacuum below $10^{-6}$ mbar. The outgassing rate measured was below $10^{-9}$ mbar/s, which allows us to avoid permanent gas circulation through special cleaning filters and to operate instead in seal mode. The gas was introduced into the vessel through a simple oxisorb filter.

Data taken at different source distances show no evidence of loss of signal intensity due to electron attachment. In Figure 6 shows the recorded energy spectrum obtained irradiating the detector with a $^{55}$Fe source, the energy resolution obtained is 18% (FWHM) in Ar + 10% $CO_2$ gas mixture.

So far the prototype has been operated with two different gas mixtures, namely Ar + 10% $CO_2$ as well as Ar + 2% Isobutane; and at different pressures up to 1.5 bar. These first tests showed that even with such a simple amplification element, high gas gains (above $10^4$) are easily achieved. Figures 7 and 8 show the obtained gain versus voltage using Argon as carrier gas mixed with 2% Isobutane and 10 % $CO_2$ respectively for various gas pressures. Notice the advantage of the Isobutane mixture requiring twice lower voltage at same pressure.

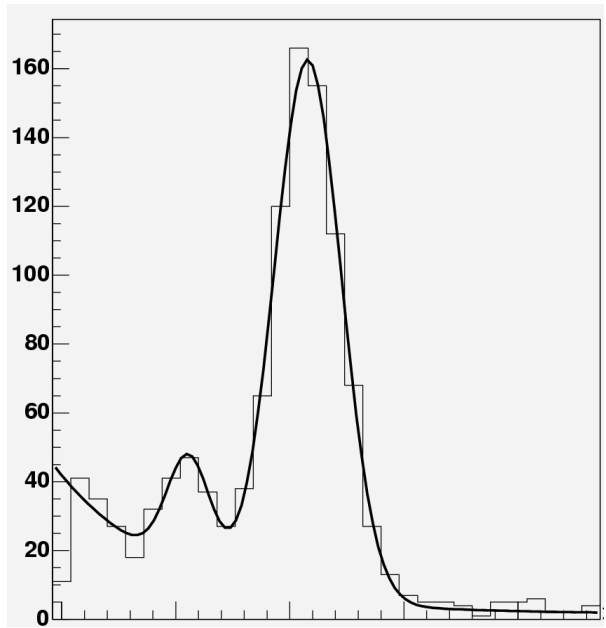

Figure 6. Pulse height distribution of the signal produced by a $^{55}$Fe source

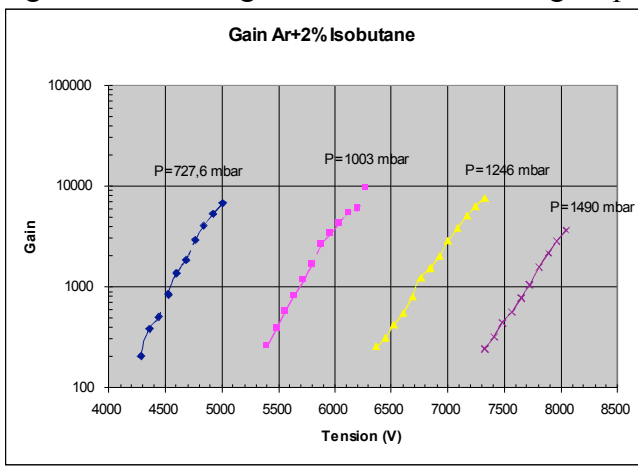

Figure 7. Effective gas gain versus high voltage for various pressures in Ar + 2% Isobutane

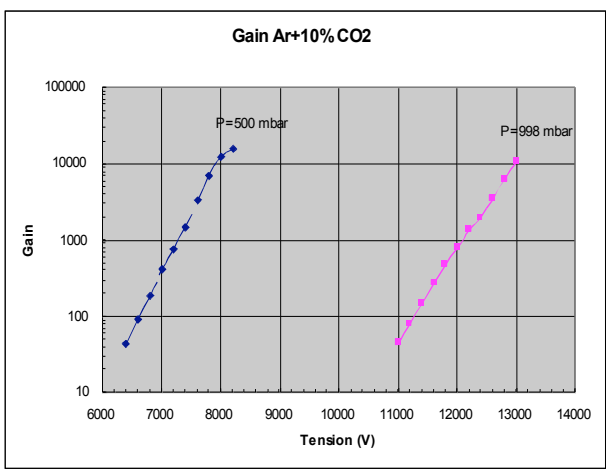

Figure 8. Effective gas gain versus high voltage for various pressures in Ar + 10% $CO_2$

Stable operation was tested up to 40 days, without gas circulation (seal mode). Runs using calibration sources of [109]Cd and [55]Fe or cosmic rays have been also performed. A typical detector response at 22 keV in Ar + 10% $CO_2$ gas, from a [109]Cd source, is shown in Figure 9.

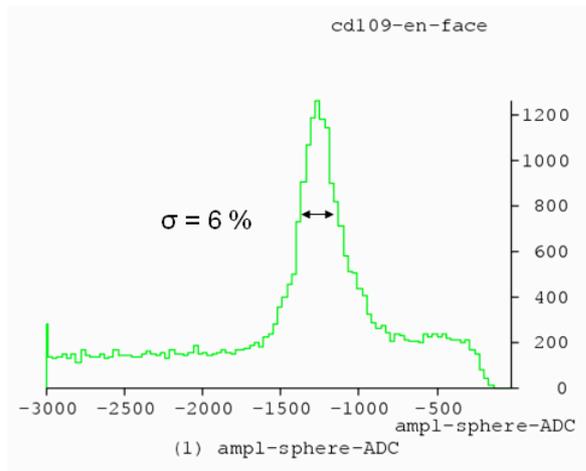

Figure 9. Pulse height distribution of the signal produced by a [109]Cd source

High gain combined with low electronic noise provides low energy threshold; this is also well illustrated in Fig.3 showing that thresholds below 100 eV are already at hand.
At low electric fields (E) the drift velocity is roughly proportional to E and the longitudinal diffusion coefficient (D) is inversely proportional to the square of E. So the time dispersion is correlated to the cube of the radius. So there is an information contained in the signal that is correlated to the radial distance of the interaction. By measuring the time dispersion of the signal with a deconvolution method one can find with a precision of about 10 cm, especially at large distances, the depth of the interaction as shown on figure 10.

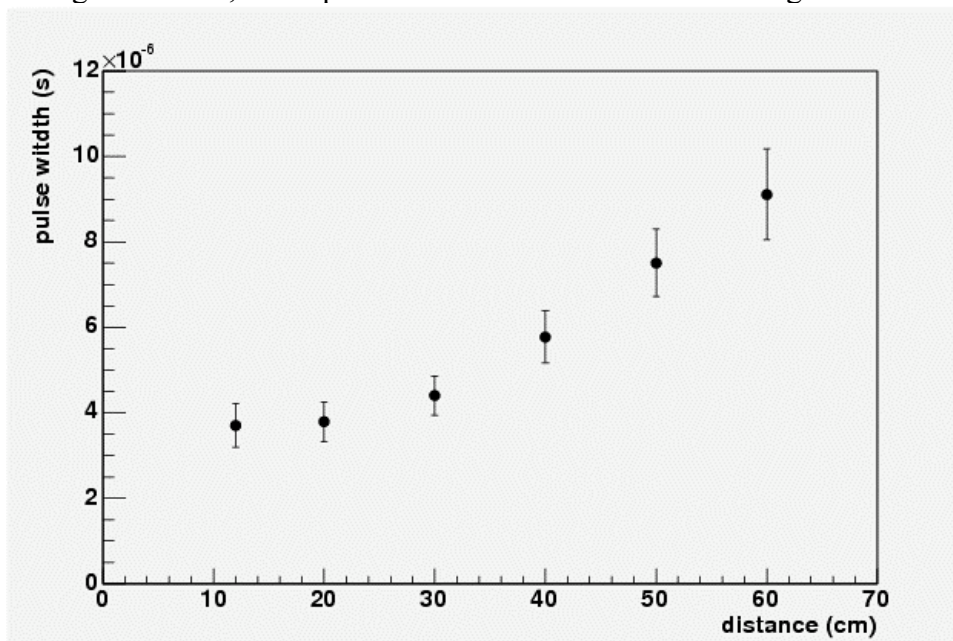

Figure 10. Time dispersion of the signal as function of the depth of interaction

## 6. Results with a field corrector

To solve the problem of field distorsion due to voltage anode electrode one needs to add one or more electrodes to the system, in order to correct the field lines towards an electrostatic configuration which:
1) approaches more an ideal spherical symmetry;
2) makes all (or most) of the field lines end up in the spherical electrode, and not in the rod; a
3) the amplification associated with each of those field lines is sufficiently similar, so the detector response is homogeneous over the whole volume. It will be shown later that 1) is probably not easy to achieve together with 2) and 3), and that achieving 2) and 3) relaxing 1) is certainly possible keeping the robustness and simplicity of the original concept, and an interesting option for many applications.

Several practical geometries have been tried to accomplish such objectives. The additional elements have been one of more of the following: annular electrodes placed along the rod at different voltages, a conical piece of resistive material placed under a voltage drop at the end of the rod or one or more concentric cylindrical electrodes placed around the rod at different voltages and at different distances from the sphere [9]. Although all tested solutions are valid in theory, the selection of the preferred one is done by practical and technical issues, in particular, the ease of construction or the absence of sparking when going to higher voltages. In this sense, the more successful configuration, (and the one with which we have obtained the results presented in the rest of the paper) is the one with a cylinder around the high voltage rod, placed at 4 mm away from the central sphere and powered with an independent voltage $V_2$ (which can be zero, i.e., at ground). The equipotential lines for the described "corrected" configuration are shown in Figure 11.

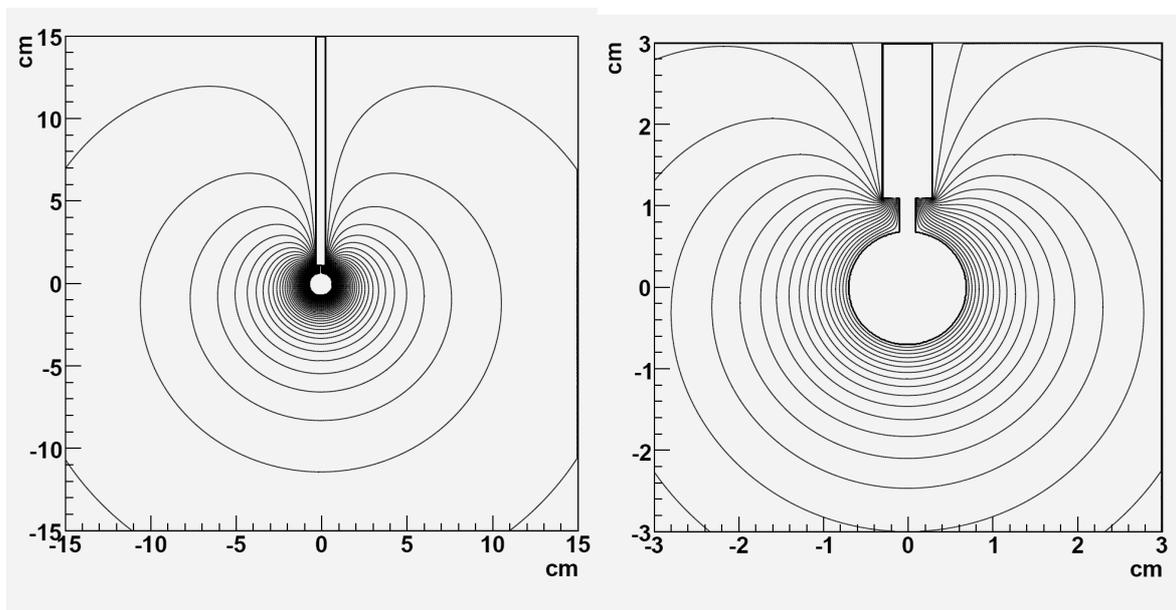

Figure 11.- Electrostatic configuration for the readout electrode, with the "corrector" electrode consisting of a grounded cylinder surrounding the high voltage rod and placed 4 mm away from the spherical electrode.

As can be seen, the presence of the cylinder at ground twists the field lines around the spherical electrode in a way that makes them more spherically symmetric than before.

Nevertheless, a closer inspection to the region close to the electrode surface still shows differences, especially near the connection to the rod. However, the relevant feature of the proposed geometry, which is responsible of its success of operation, is illustrated in Figure 12, where the corresponding field lines are drawn. As can be seen, the presence of the cylinder at ground pushes the field lines away, focusing most of them in a "constrained" region of the spherical electrode, the one opposite the rod (where the amplification happens to be more homogeneous). What could be seen as an unpleasant feature, away from the spherical symmetry, is however a positive one, as the amplification suffered along electron path is rather similar making the detector response remarkably homogeneous for most of the detector volume.

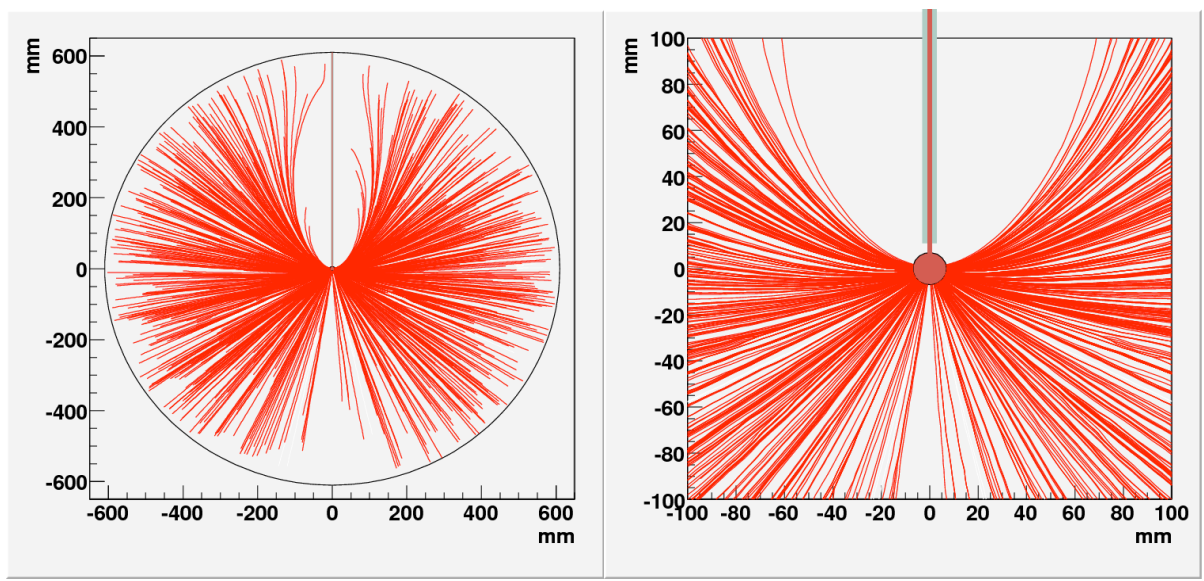

**Figure 12.- Drift field lines for the geomtry with the cylinder at ground. The lines have been calculated starting form 1000 points randomly distributed over the sphere volume. In this case all lines end up on the spherical electrode, and are pushed away from the conflictive region of the connection sphere-stick. Left one shows the overall picutre and right a close-up of the 10 cms around the center.**

These facts have been proved by the experimental measurement of the energy resolution for a gaseous source (radon) homogeneously active over the full detector volume, as shown in Figure 13; a good energy resolution of .5% has been achieved. A dedicated paper on the high resolution investigation is in preparation. The remarkable results achieved show that this extremely simple and robust readout configuration is a very appealing working option for cases where perfect radial drift lines are not required.

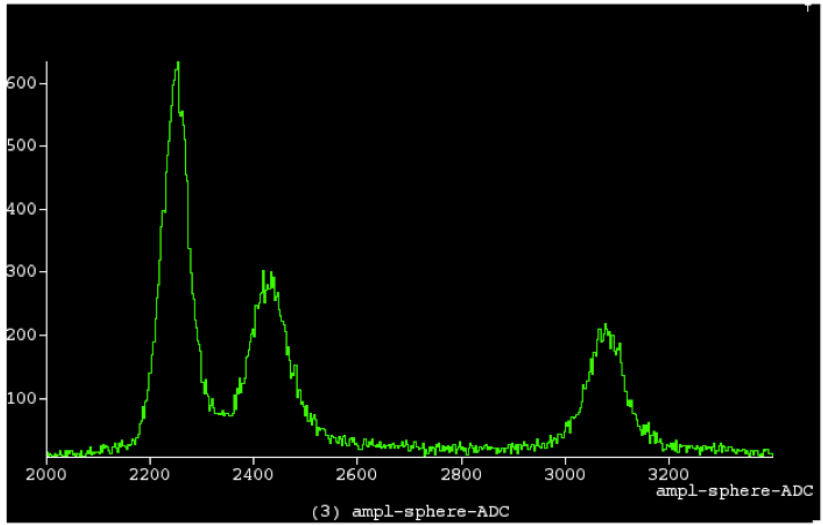

Figure 13. Peaks observed from a $^{222}$Rn radioactive source. From left to right we observe the $^{222}$Rn peak at 5 MeV, the $^{218}$Po and $^{214}$Po at 6 MeV and 7.7 MeV respectively

**7. Future developments and applications**
More complicated readout geometries have being studied and other ideas will be developed. An evident option is the addition of one or more cylinders concentrical to the one above described. This would introduce intermediate steps between ground and the high voltage of the central electrode, improving the sphericity of the drift and reducing the voltage between electrodes allowing for higher working voltages/gains. These improvements are at the cost of course of reducing the simplicity of the present design, adding additional powered electrodes. Another very appealing option, although at a more complex level, is the one in which the spherical electrode is replaced by a small Micromegas structure [14-17]. Its main advantage is that the amplification and drift fields are decoupled, and the geometrical requirements are therefore less stringent. In addition, the amplification in a Micromegas is well known and has well-proven nice features in terms of energy and time resolution [18-21]. Results with the present spherical detector equipped with a Micromegas readout are left for a forthcoming publication
Further developments are needed to optimize the detector for high pressure operation and especially with Xenon gas mixture. High pressure Xenon gas fillings, together with good energy resolutions, extend the detection range to hard X-rays and generally to gamma spectroscopy. Good energy resolution is crucial for future double decay studies using enriched $^{136}$Xe target in order to separate neutrinoless double beta decay (0νββ mode) from allowed double decay accompanied by two neutrinos (2νββ mode). The 0νββ process is expected to occur only if neutrinos are Majorana particles and turns to be the most sensitive process for measuring a small neutrino mass.
Adding a small fraction of $^3$He gas this detector is ideally suited for neutron detection and neutron energy measurement. The counter can be applied in experiments requiring low background and a low energy threshold. An example is the detection of the coherent neutrino-nucleus scattering producing sub keV ion recoils. A spherical detector of radius 4 m and employing Xe gas at a pressure of 10 Atm will detect about 1000 events for a typical supernova explosion at 10 kpc. A world wide network of several such simple, stable and low cost supernova detectors is proposed [22].

## 8. Conclusions

We have developed a new detector based on the radial geometry with spherical proportional amplification read-out. The main advantage of this structure is the use of a single electronic channel to read-out a large volume. This single information still allows the determination of the radial coordinate of the interaction point through the measurement of the time dispersion of the detected charge pulse. Such information is of paramount importance for localization in depth and background rejection applying fiducial cuts.

It is the most cost-effective way of instrumenting a large detector volume with a minimum of front-end electronics. This approach simplifies the construction and reduces the cost of the project. Large gains were obtained providing low energy threshold in the sub-keV region.

A simple field corrector has been developed and good energy resolution has been measured over all the volume of the spherical vessel. The detector is robust, stable in time and operates in seal mode.

Several applications are open arising from low energy neutrino physics, double beta decay to WIMP search, Supernova detection or neutron background measurement.